\documentclass[aps,prb,twocolumn,superscriptaddress,floatfix,10pt]{revtex4-1}

\usepackage{graphicx}
\usepackage{amsmath}
\usepackage{epsfig}
\usepackage{helvet}
\usepackage{amssymb}
\usepackage{algorithm}
\usepackage{algpseudocode} 

\usepackage{enumitem}
\usepackage[bookmarksnumbered=true,hypertexnames=false,colorlinks=true,linkcolor=blue,urlcolor=blue,citecolor=blue,anchorcolor=green]{hyperref}
\usepackage{framed}
\usepackage{bbm}
\usepackage{dsfont}

\newcommand{\ket}[1]{\mbox{$| #1 \rangle$}}

\newcommand{\frw}[1]{$\overset{\lower0.5em\hbox{$\smash{\scriptscriptstyle\smile}$}} #1$}

\begin{document}

\title{Reconstruction of Randomly Sampled Quantum Wavefunctions using Tensor Methods}
\author{Aaron Stahl}
\author{Glen Evenbly}
\affiliation{School of Physics, Georgia Institute of Technology, 830 State Street, Atlanta, GA 30332 }
\email{astahl3@gatech.edu}
\date{\today}

\begin{abstract}
We propose and test several tensor network based algorithms for reconstructing the ground state of an (unknown) local Hamiltonian starting from a random sample of the wavefunction amplitudes. These algorithms, which are based on completing a wavefunction by minimizing the block R{\'e}nyi entanglement entropy averaged over all local blocks, are numerically demonstrated to reliably reconstruct ground states of local Hamiltonians on 1D lattices to high fidelity, often at the limit of double-precision numerics, while potentially starting from a random sample of only a few percent of the total wavefunction amplitudes.
\end{abstract}

\maketitle

\section{Introduction}
Tensor network methods \cite{TN1,TN2,TN3} are integral to the study of quantum many-body systems and have also found a wide range of applications in areas such as quantum chemistry\cite{QC1,QC2}, machine learning\cite{ML1,ML2,ML3}, and holography\cite{Holo1,Holo2,Holo3,Holo4}. In the context of quantum many-body wavefunctions, the exponential growth of the Hilbert space in system size makes exact simulation of these systems computationally intractable at scale. Tensor networks can potentially bypass this obstacle by allowing wavefunctions to be expressed efficiently as a product of many small tensors. Various algorithms have been proposed that utilize this compressed form, including White's density matrix renormalization group \cite{DMRG1,DMRG2,DMRG3} and its pioneering use of matrix product states \cite{MPS1,MPS2,MPS3} (MPS), projected entangled pair states \cite{PEPS1,PEPS2,PEPS3,PEPS4}, tree tensor networks \cite{TTN1, TTN2}, and the multi-scale entanglement renormalization ansatz \cite{MERA1,MERA2,MERA3,MERA4}. Many of these tools allow properties of the ground states of quantum many-body body systems to be accurately computed, even in the thermodynamic limit.

Key to the success of tensor network methods is that the ground states of quantum many-body systems, which obey an area law for entanglement scaling \cite{Ent1,Ent2,Ent3,Ent4,Ent5,Ent6}, are highly atypical as compared with a random state of the Hilbert space. The goal of the present manuscript is to leverage this atypicality, utilizing MPS and tree tensor networks, to accurately reconstruct quantum many-body wavefunctions starting from only a (randomly sampled) subset of the total wavefunction amplitudes. Application of tensor network methods to the task of wavefunction reconstruction, starting from only a partial sample, represents a novel use case for tensor networks and opens up potential new avenues of study across a broad range of disciplines in physics and data science.

The idea of using a small subset of randomly sampled amplitudes to solve for the complete wavefunction is an example of a much more general problem known as matrix and tensor completion. In completion problems, the principle goal is to reconstruct (or ``complete") a data set using a subset of data points from the desired set. While work on matrix completion dates back to the 1990s, interest in matrix completion burgeoned in the mid-2000s, with Cand{\`e}s, Recht, and others making great strides in both algorithm development and theoretical findings \cite{MCM1,MCM2,MCM3,MCM4,MCM5,MCM6,MCM7,MCM8,MCM9,MCM10,MCM11}. Matrix completion has proven to be exceptionally useful in a broad range of applications including recommender systems \cite{MCA10}, data analysis \cite{MCA1, MCA2}, bioinformatics \cite{MCA3,MCA4}, imaging sciences \cite{MCA5,MCA6}, compressed sensing \cite{Sens1,Sens2,Sens3}, and other areas. Ideas from matrix completion have also been generalized to higher-dimension data structures yielding \emph{tensor completion}. Tensor completion has also seen growing interest, with numerous theoretical and algorithmic developments occurring in past several years \cite{TCM1,TCM2,TCM3,TCM4,TCM5}. It has found broad application in areas such as big data \cite{TCA1}, pattern analysis \cite{TCA2}, and hyperspectral image recovery \cite{TCA3}.

In this work we explore the application of tensor completion to the reconstruction of quantum many-body wavefunctions, focusing mainly on ground states of various 1D spin chains. Ground state wavefunctions are found to possess properties that pose difficulty 
for established matrix and tensor completion algorithms, such that they perform poorly in this setting. Accordingly, we propose and test several new approaches to tensor completion using tensor networks, which are demonstrated to significantly outperform the established algorithms when applied to wavefunction reconstruction.

\section{Wavefunction completion: definition, scope, notation}
In this section, we strictly define wavefunction completion and establish the scope of many-body quantum wavefunctions considered in this study. We limit our study to systems that can be expressed as one-dimensional (1D) finite lattices of $N$ sites, with local dimension $d$ per lattice site, and two-body interactions that couple sites up to $l$ distance from one another. Mathematically, this encompasses all Hamiltonians that satisfy the form,
\begin{equation}
    H |\Psi \rangle = \sum_{i=1}^{N-(l-1)} h_{i,i+l} \ket \Psi , 
\label{eq:def_ham}
\end{equation}
where $h$ is a local operator coupling the $i^{\text{th}}$ site to the $(i+l)^{\text{th}}$ site. (\ref{eq:def_ham}) corresponds to a system with open boundary conditions but the expression is easily extended to systems with periodic boundary conditions. Treatment of both conditions are included in our analysis. The Hilbert space encompassed by (\ref{eq:def_ham}) is relatively expansive and includes a rich array of physical systems with diverse singular value spectra, from highly entangled critical systems like the XX-model to zero entanglement product states like the Ising model with no magnetic field ($\lambda=0$).

Consider next the eigenstates of such Hamiltonians, $\ket \Psi$; if we let $\{ |j_{1} \rangle \cdots |j_{d} \rangle \}$ be an orthonormal basis for a Hamiltonian on a 1D finite lattice with $N$ sites (eg., \{$\ket{0},\ket{1}\}$ forms an orthonormal basis for a single spin system where $d=2$), we can represent an eigenstate of the system as\cite{MPS3},
\begin{equation}
    |\Psi \rangle = \sum_{i_{1}=0}^1 \cdots \sum_{i_{n}=0}^1 c_{i_{1} \cdots i_{N}} |i_{1}\rangle \otimes \cdots \otimes |i_{N} \rangle ,
\label{eq:def_psi}
\end{equation}
where the number of coefficients $\{c_{i_1,...,i_N}\}$ required to specify the state grows exponentially in system size as $d^N$. The problem of wavefunction completion can now be defined as follows: Given a random partial sample of coefficients, $\{c_1,...,c_{k}\} \in S$, where $S$ is a set of randomly sampled coefficients and $k<d^N$, as well as the locations within the state $\vec{s}$ of those $k$ coefficients, estimate the values of the ($d^N-k$) unsampled coefficients in the complement $S'$. Estimating the values of the unsampled coefficients in $S'$ using the sampled coefficients in $S$ constitutes ``completing" the state. This is exactly analogous to a typical matrix or tensor completion problem. However, whereas typical data sets of interest in matrix and tensor completion correspond to structures with fixed dimensionality, we can potentially reshape wavefunctions to evaluate different bipartitions of the state. This aspect of wavefunctions is explored further in the section on algorithm development (see Sec. IV). To evaluate the accuracy of a completed state, we use the fidelity $f$ of the result,
\begin{equation}
    f(\ket \Psi,\ket{\Psi_{C}}) = \frac{\langle \Psi | \Psi_{C} \rangle \langle \Psi_{C} | \Psi \rangle}{\langle \Psi | \Psi \rangle \langle \Psi_{C} | \Psi_{C} \rangle},
\label{eq:def_fidelity}
\end{equation}
where $\ket{\Psi}$ is the exact wavefunction and $\ket{\Psi_{C}}$ is the completed wavefunction. The associated \textit{fidelity error} is then $\varepsilon = 1 - f$. Using this terminology, we can define wavefunction completion as the following optimization problem,
\begin{equation}
\begin{split}
    & \text{minimize}\; \varepsilon \quad \text{w.r.t.} \quad \{c'_1,...,c'_{d^N-k}\} \in S' \\
    & \text{such that} \quad \ket{\Psi_{C}(\vec{s})} = \ket{\Psi(\vec{s})}
\label{eq:def_problem}
\end{split}
\end{equation}

Using the notation above, we also introduce construction of the \textit{incomplete} wavefunction passed as the input to a matrix or tensor completion algorithm. First consider the initial guess given to entries of $S'$. It was proven by Ge et al. \cite{MCM9} that positive semi-definite matrix completion does not depend the initial guess assigned to unobserved/unsampled entries. Locally defined ground states satisfying (\ref{eq:def_ham}) are not typically positive semi-definite about any bipartition, but in practice we observe that the initial guess of $S'$ has negligible impact on the final result of completion, barring certain limited exceptions\footnote{Setting the initial guess to a local minima can result in failed convergence to a global minimum}. Specifically, we found an initial guess of zero, sampling from distributions consistent with coefficients in $S$, and insertion of random noise spanning many orders of magnitude all produced approximately the same convergence properties. For the sake of simplicity, we thus set the initial guess of all unsampled coefficients $S'$ to zero. The wavefunction completion algorithm input can therefore be expressed as follows,
\begin{equation}
    \ket{\Psi_{C_0}} = \ket{\Psi(\vec{s})}+\ket{\Psi(\vec{s'})}*\vec{0},
\label{eq:def_input}
\end{equation}
where $\ket{\Psi_{C_0}}$ is the input and includes all coefficients of the wavefunction to be completed, with sampled coefficients fixed to their true values and unsampled coefficients set to zero. We leave room for a more sophisticated approach to initial guesses that may result in faster convergence.

\section{Theoretical motivations and initial results}

\begin{figure} [!h] 
\begin{center}
\includegraphics[width=8.5cm]{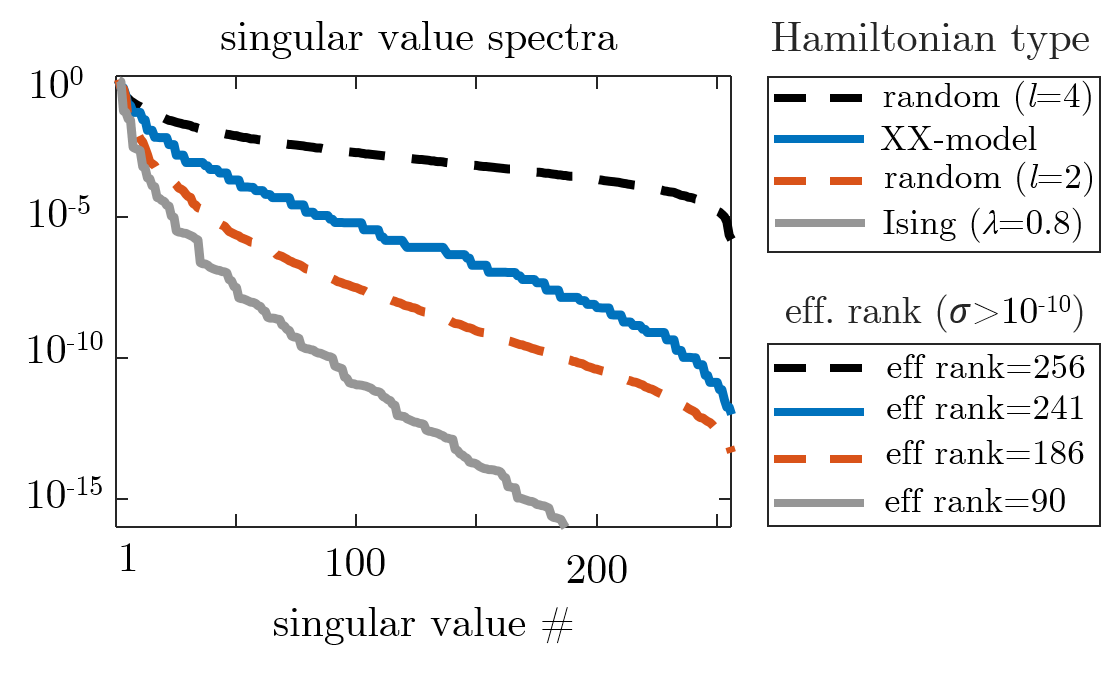}
\caption{Singular value spectra for local ground states on a $N=16$ site lattice with periodic boundary conditions, where each line is labeled in the upper right. Both of the random ground states (dashed lines) correspond to randomly generated inhomogeneous complex Hamiltonians with local dimension $d=2$, varying only in the interaction length $l=4$ (above, black) and $l=2$ (below, red). Ground states for the XX model and Ising model with magnetic field strength parameter $\lambda=0.8$ are also included (solid lines). The effective rank of each corresponding ground state, given by the cutoff $\sigma > 1\text{e-}10$, are provided in the lower right box for reference, where maximum rank for a 16-site lattice of local dimension $d=2$ is 256.}
\label{fig:svals_example}
\end{center}
\end{figure}

Conventional matrix and tensor completion rely on the fundamental restriction that the matrix or tensor of interest be low rank. If there is no rank restriction, the problem of completing that matrix or tensor is ill-posed, as the unknown entries could be assigned any arbitrary value. This presents a potential problem for wavefunctions completion, as wavefunctions have no inherent rank restriction and even locally defined ground states, which are known to have limited entanglement enforced by area law restrictions, often contain non-negligible information buried deep in the singular value spectra. Unlike the spectra encountered in typical matrix or tensor completion problems, which fall abruptly to zero in the low-rank regime, locally defined ground states are characterized by exponentially decaying spectra in which approximately all singular values are needed for an exact completion.

In Fig. \ref{fig:svals_example}, a range of spectra for locally defined ground states are plotted. Of note is that even with a minimal nontrivial interaction length $l=2$ on a $N=16$ site lattice, the effective rank ($\sigma > 1\text{e-}10$) $r_{\text{eff}}=186$ out of a possible maximum effective rank of 256, and never goes to zero. This means that an exact completion is strictly not possible. For example, even with one missing complex coefficient, where the magnitude is set by the normalization constraint, the corresponding phase could hypothetically take on any arbitrary value. When interaction length is increased to $l=4$ in Fig. \ref{fig:svals_example}, the resulting ground state is full-rank by several orders of magnitude under the same effective rank condition ($\sigma_{256} \approx 1{\text{e-}}6$ vs. $r_{\text{eff}}=1{\text{e-}}10$). Fig. \ref{fig:svals_example} also includes two well-studied physical models, the critical XX-model and the Ising model with a sub-critical magnetic field strength parameter $\lambda = 0.8$ (as $\lambda \rightarrow 0$, the Ising model becomes an approximate product state, with no entanglement and $\approx 1$ non-zero singular values; e.g., for $N=16$, $\lambda=0$ yields 2 non-zero singular values). The XX-model in particular shows how a highly entangled critical system has singular values outside the low-rank regime that are several orders of magnitude larger than the equivalent in a sub-critical or randomly generated system of the same interaction length $l=2$. Overall, these spectra profiles highlight the challenge posed by wavefunction completion, even in ground states, and underscore the importance of integrating highly numbered singular values.

\begin{figure} [!t!b]
\begin{center}
\includegraphics[width=8.5cm]{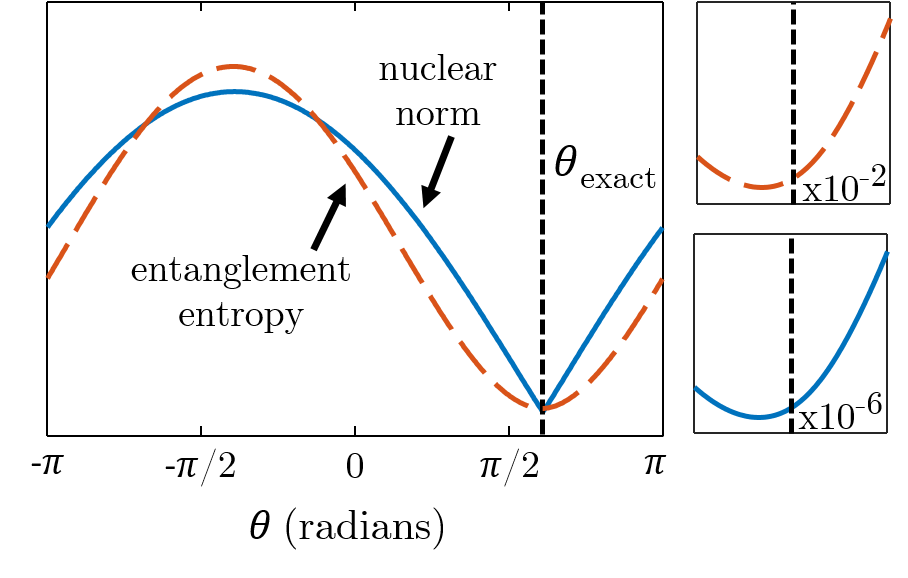}
\caption{Sweep of entire parameter space for the phase of a randomly chosen coefficient of a local ground state. The state is defined by a randomly generated inhomogeneous complex Hamiltonian with local dimension $d=2$ and interaction length $l=2$ on a $N=14$ site lattice with periodic boundary conditions. Both the $S_{1/2}$ entanglement (solid blue line) and $S_{1}$ entanglement (dashed red line) are included, with the exact phase $\theta_{\text{exact}}$ indicated by the black dashed vertical line. The figures on the right are zoomed about the minimum $S_{1/2}$ (lower-right) and $S_{1}$ (upper-right) $\theta$-values and list the $\theta$ range spanned by the x-axes, indicating the respective proximity to $\theta_{\text{exact}}$.}
\label{fig:phase_sweep}
\end{center}
\end{figure}

Given that locally defined ground states are often high/full-rank, rank minimization for ground state completion is not a feasible solution even in theory for wavefunction completion. Rank minimization is also problematic in the context of matrix/tensor completion due to its general impracticality, being NP-hard and thus computationally intractable in most settings\cite{MCM8}. Researchers in low-rank matrix completion found that minimizing the nuclear norm, or trace over the singular values, offers an excellent and practical proxy for minimizing the rank. This likewise affords a natural starting point for the problem of wavefunction completion.

Throughout this paper, we refer to the entanglement of a ground state and take this to mean the degree of quantum entanglement between two subsystems of a state. In the context of many-body quantum systems, this commonly refers to the bipartite Von Neumann entanglement entropy or the R{\'e}yni entropy of order 1. Mathematically, this entanglement can be expressed as, $S_{1} = -\sum_{k} |\sigma_{k}|^2 \text{log}(|\sigma_{k}|^2)$, with $\sigma_{k}$ being the $k^{\text{th}}$ singular value across a particular bipartition. The nuclear norm is another measure of quantum entanglement and often arises in data science and related fields. It is equivalent to the R{\'e}nyi entropy of order $\frac{1}{2}$ and can be expressed as $S_{1/2} = \sum_{k} \sigma_{k}$, or the trace over the singular values. We refer to these two measures of entanglement as the $S_{1}$ entanglement and $S_{1/2}$ entanglement and unless otherwise specified, use the center-most bipartition when directly measuring either quantity. 

\begin{figure} [!t!b]
\begin{center}
\includegraphics[width=9.0cm]{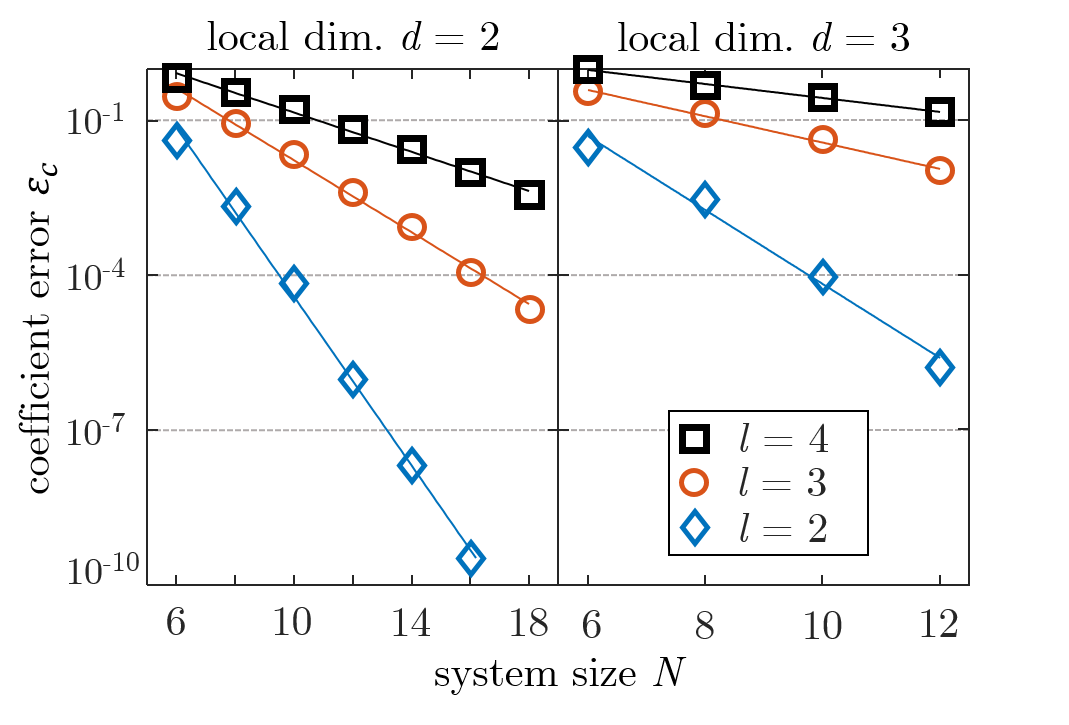}
\caption{Fitting the coefficient error $\varepsilon_{c}$ as a function of system size $N$ to an exponential curve, $\varepsilon_{c} \approx \exp(-\alpha N)$, for systems with local dimension $d=2$ (left) and $d=3$ (right) at varied interaction length $l$. Fitting coefficient values can be found in the associated  Table \ref{table:alpha_table}. Coefficient error $\varepsilon_{c}$ is defined as the absolute difference between the coefficient value $c_{1/2}$ that minimizes the $S_{1/2}$ entanglement, which is calculated directly for each coefficient, and the exact coefficient value $c$, or $\varepsilon_{c} = |c-c_{1/2}|$. Each data point is normalized for system size by dividing by $d^N$ prior to fitting. To generate the results, we used 50 ground states of 50 different randomly generated inhomogeneous complex Hamiltonians with periodic boundary conditions at each $N$, $d$, and $l$, and found the median $\varepsilon_{c}$ in all cases. We then took the median of those 50 median coefficient errors to obtain each fitting point.}
\label{fig:alpha_plot}
\end{center}
\end{figure}

Studying the relationship between ground state coefficients and entanglement can help inform whether $S_{1}$ and/or $S_{1/2}$ minimization offers an effective potential approach to wavefunction completion. To that end, we consider the ``completion" of the phase of a randomly selected coefficient of a locally defined complex ground state. This constitutes the smallest nontrivial completion problem for a complex wavefunction, wherein all entries are randomly sampled until a single coefficient remains. Using the normalization constraint, $\langle \Psi | \Psi \rangle = 1$, to fix the magnitude of this coefficient, we are left with just the unknown phase. Establishing how the exact phase of the coefficient $\theta_{\text{exact}}$ relates to phase(s) that minimize the $S_{1}$ and $S_{1/2}$ entanglement, $\theta_{S_{1/2}}$ and $\theta_{S_{1}}$, respectively, will aid our understanding of whether or not entanglement minimization is an effective approach to ground state wavefunction completion. To find the minimal entanglement values, the nuclear norm ($S_{1/2}$) and entanglement entropy ($S_{1}$) are both independently minimized across the central bipartition over the entire parameter phase space using the Nelder-Mead simplex algorithm as described in Lagarias et al. \cite{MISC1}. 

\begin{table} [!t!b] 
\begin{center}
\includegraphics[width=8.0cm]{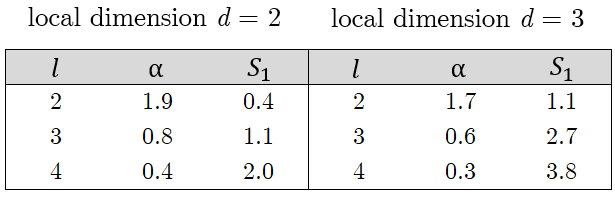}
\caption{Tabled data corresponding to Fig. \ref{fig:alpha_plot}, where $l$ is the interaction length, $\alpha$ is the exponential fitting coefficient, $d$ is the local dimension, and $S_{1}$ is the R{\'e}nyi entanglement entropy of order 1. $S_{1}$ is taken across the central bipartition at $N=10$ for each type of system. $S_{1}$ is generally stable across the systems sizes included in our fits, and varies by less than 5\% from $N=6$ to $N=18$ in the $d=2$ case and similarly for $d=3$.} 
\label{table:alpha_table}
\end{center}
\end{table}

Fig. \ref{fig:phase_sweep} plots the $S_{1/2}$ and $S_{1}$ entanglement as a function of the phase of the unsampled coefficient for the ground state of a randomly generated complex inhomogeneous Hamiltonian on a 1D 14-site lattice of local dimension $d=2$ and interaction length $l=3$. The result highlights the typical relationship between $\theta_{\text{exact}}$ and the two minimum entanglement values, $\theta_{S_{1/2}}$ and $\theta_{S_{1}}$. Both minimums offer reasonable approximations to the exact phase and as we will see later, the differences between $\theta_{S_{1/2}}$, $\theta_{S_{1}}$, and $\theta_{\text{exact}}$ approach zero as system size $N$ grows arbitrarily large. However, for practical purposes we are computationally limited to completion of system sizes $N \leq 20$, and in this regime the phase that minimizes the nuclear norm $S_{1/2}$ is several orders of magnitude closer to exact phase. We will refer to the set of values that minimize the nuclear norm as the minimal $S_{1/2}$ entanglement completion moving forward.

To further understand and quantify the relationship between the complex coefficient values that minimize the $S_{1/2}$ entanglement and key parameters that define a local ground state (i.e., system size $N$, local dimension $d$, and interaction length $l$), we consider the completion of ground states in which all but two complex coefficients are unsampled. In other words, $(d^N-2)$ coefficients are randomly sampled, leaving only two coefficients in $S'$. Because the coefficients are complex, there are initially four free parameters (two magnitudes and two phases). However, due to the wavefunction normalization constraint, the two magnitudes are coupled and only represent one free parameter. Using this configuration, we exactly minimize the $S_{1/2}$ entanglement over the full parameters space and consider the absolute error $\varepsilon_{c} = |c-c_{1/2}|$ of the coefficient for which the magnitude is allowed to vary, letting $c$ be the exact coefficient value and $c_{1/2}$ the complex value that minimizes the $S_{1/2}$ entanglement. After single phase completion as performed in Fig. \ref{fig:phase_sweep}, this is the second ``easiest" nontrivial completion problem for complex wavefunctions.

Varying $N$, $d$, and $l$, and generating 50 different random ground states at each combination thereof, we fit the resulting median coefficient error $\varepsilon_c$ to an exponential in system size, $\varepsilon_{c} \approx \beta \; \text{exp}(-\alpha N)$, holding $d$ and $l$ constant in each fit. The results are displayed in Fig. \ref{fig:alpha_plot} and Table \ref{table:alpha_table}. Note the $\beta$ values aren't of interest in characterizing the scaling behavior in $N$, so we don't elaborate further on their significance. As seen in Fig. \ref{fig:alpha_plot}, the coefficient errors are well approximated by an exponential in $N$, indicating that $\varepsilon_c$ goes to zero exponentially quickly in $N$. In other words, as $N$ grows large, the minimal $S_{1/2}$ entanglement completion offers an increasingly accurate representation of the exact wavefunction. This in and of itself is an interesting result; given the limitations imposed by the 1D area law on locally defined ground states, it is expected that entanglement across any particular bipartition would be small. However, a priori, it was not obvious that the value that minimizes the entanglement would precisely match the exact value of the coefficient if $N$ is sufficiently large. As expected, we also find that the rate $\alpha$ at which $\varepsilon_{c}$ goes to zero is inversely related to the local dimension $d$ and interaction length $l$, both of which contribute to entanglement at fixed $N$. This is also confirmed by the inverse relationship between $\alpha$ and the entanglement entropy $S_{1}$ listed in Table \ref{table:alpha_table}.

This result confers theoretical maximum accuracy bounds on the minimal $S_{1/2}$ entanglement completion of a local ground state. In other words, any algorithm that successfully finds the minimum nuclear norm with respect to a partial sample of coefficients will have its accuracy fundamentally limited by the $\alpha$ value corresponding to that system. This deviates from established results in low-rank matrix and tensor completion, where as long as you have a sufficient number of entries in a partial sampling, the matrix or tensor can be completed with exact precision. With locally defined ground states, it can be the case that no sample rate is sufficient to achieve a near-perfect completion. However, despite not being able to exactly complete some ground states, our fitting results demonstrate that as long as a system is relatively large compared to its local dimension and interaction length, it can potentially be completed to a high degree of accuracy vis-{\`a}-vis obtaining the minimal $S_{1/2}$ entanglement solution. 

While exactly minimizing a set of unsampled coefficients is feasible when the cardinality is small, this approach quickly becomes intractable as the number of unsampled coefficients increases to hundreds, thousands, or potentially millions. Hence, we are motivated to construct an algorithm that reliably finds the minimal $S_{1/2}$ entanglement solution when large numbers of coefficients are unsampled.

\section{Algorithm overview}

Recall the wavefunction completion problem: using a random partial sample of a wavefunction's coefficients $S$ and their corresponding locations within the state $\vec{s}$, estimate the unsampled coefficients in $S'$ to minimize the fidelity error $\varepsilon$, as stated in \ref{eq:def_problem}. In doing so, assume no knowledge of rank, system size, Hamiltonian type, local dimension, and interaction length. Whereas minimizing the $S_{1/2}$ entanglement across a single bipartition was sufficient for the simplified completion problem in Sec. III, where we exactly minimized the nuclear norm, this is not generally the case. Indeed, wavefunction completion across a single bipartition constitutes a matrix completion problem and there exist many mature and effective algorithms in this domain. However, these algorithms frequently converge to a local minimum nuclear norm, where the smaller the sample rate, the more likely this is to occur. In these cases, there exist possible completion solutions with lower nuclear norms about the matricized bipartition, but without leveraging other partitions of the wavefunction they remain inaccessible to fixed-dimension (e.g., matrix) completion routines. Furthermore, even within the context of fixing a wavefunction about a particular bipartition, 

Overall, we attribute the sub-optimal performance of existing matrix and tensor completion algorithms to the exponential decay pattern and high-rank properties of wavefunction spectra, as well as the fixed treatment of dimensionality inherent to contemporary matrix and tensor completion methods. This motivates the development of our own algorithm for wavefunction completion, which must accommodate the unique set of challenges posed by of quantum states: (a) no rank restriction, (b) exponentially decaying singular value spectra, and (c) multiple physically meaningful bipartitions.

To elaborate on (c) - most incomplete data sets of interest to data scientists have fixed dimensionality (e.g., the Netflix problem is a 2-D matrix and has no additional meaning when reshaped along arbitrary dimensions). Wavefunctions, on the other hand, have physical meaning associated with every possible reshaping of the state. For example, if we consider a lattice of $N=10$ sites with local dimension $d=2$, an associated ground state would contain $d^N=1024$ coefficients. If we reshape the resulting state into a $d^n$ by $d^{N-n}$ matrix (e.g., a 16 x 64 matrix for $n=4$), we are in effect evaluating partition between the $n$ lattice sites on the left and the $N-n$ lattice sites on the right (assuming we arbitrarily view the 1D lattice sites horizontally, from left to right). This special property of wavefunctions can potentially be leveraged in a completion algorithm by making use of all nontrivial bipartitions of the state, and in particular, by minimizing the entanglement across each bipartition simultaneously. Iterating over different bipartitions offers the possibility of overcoming the local minima that commonly ensnare existing completion routines.

With these considerations in mind, we propose a sequential ramping up of bond dimension $\chi$ to accommodate the high-rank and exponential spectra decay of ground states ((a) and (b) above, respectively). Iterations are performed until the state is sufficiently converged (or 'completed') in each $\chi$, after which the $\chi$-completed state is passed as the starting point to the next bond dimension ($\chi+1$). At each $\chi$, we attempt to minimize the $S_{1/2}$ entanglement across all bipartitions simultaneously in order to fully exploit the multi-way structure inherent to wavefunctions ((c) above). This simple structure is outlined as follows, where the initial input $\ket{\Psi_{C_0}}$ at $\chi=d$, is described by (\ref{eq:def_input}) (see Sec. II),
\vspace{2.0mm}
\begin{algorithmic}
\For {$\chi = d:\chi_{\text{max}}$}
    \For {$k = 1:k_{\text{max}}$}
        \State {$\ket{\Psi_{C_k}(\vec{s'})} \leftarrow \; \text{update}\; (\ket{\Psi_{C_{k-1}}(\vec{s'})}$}
    \EndFor
    \State {$\ket{\Psi_{C_0}} \leftarrow \ket{\Psi_{C_{k_{\text{max}}}}}$} 
\EndFor .
\end{algorithmic} 
\vspace{2.5mm}
That $\chi$ begins at the smallest nontrivial value, $\chi = d$, is noteworthy, as starting even from $\chi = d+1$ often results in a failed convergence. This follows intuitively from the idea that even though ground states are often high-rank, the exponentially decaying spectra ensure that most of the information is contained in the lower bond dimensions. Neglecting any particular low $\chi$ can result in missing information that is required for the minimal entanglement completion.

As seen in the algorithmic scaffolding, the crux of the approach resides in the update step, where values of the unsampled coefficients are updated via the repeated application of truncated singular value decompositions (T-SVDs) across various bipartitions of the state. 

\section{Results}

In this section, we validate the performance of our new tensor completion algorithm and numerically demonstrate several theoretical conjectures relating the completability of local ground states to system size $N$, local dimension $d$, interaction length $l$, and entanglement entropy $S_{1}$. We also compare our tensor-based approach to established matrix completion algorithms and show that by leveraging decompositions across all possible bipartitions of a state, our algorithm exceeds the accuracy results achieved by matrix-based approaches by several orders of magnitude. For conciseness, we limit the parameter space of our results to locally defined, inhomogeneous, complex ground states with periodic boundary conditions on one-dimensional lattices unless otherwise specified. The conclusions obtained from this parameter space, however, hold across the Hilbert space of all 1D local ground states (e.g., real systems, systems with open boundary conditions, critical and sub-critical systems, systems with homogeneous local operators, and any combination thereof). Finally, we explore the limitations of completion in the context of local ground states and offer suggestions for future works.

\begin{figure}[!t]
\begin{center}
\includegraphics[width=8.5cm]{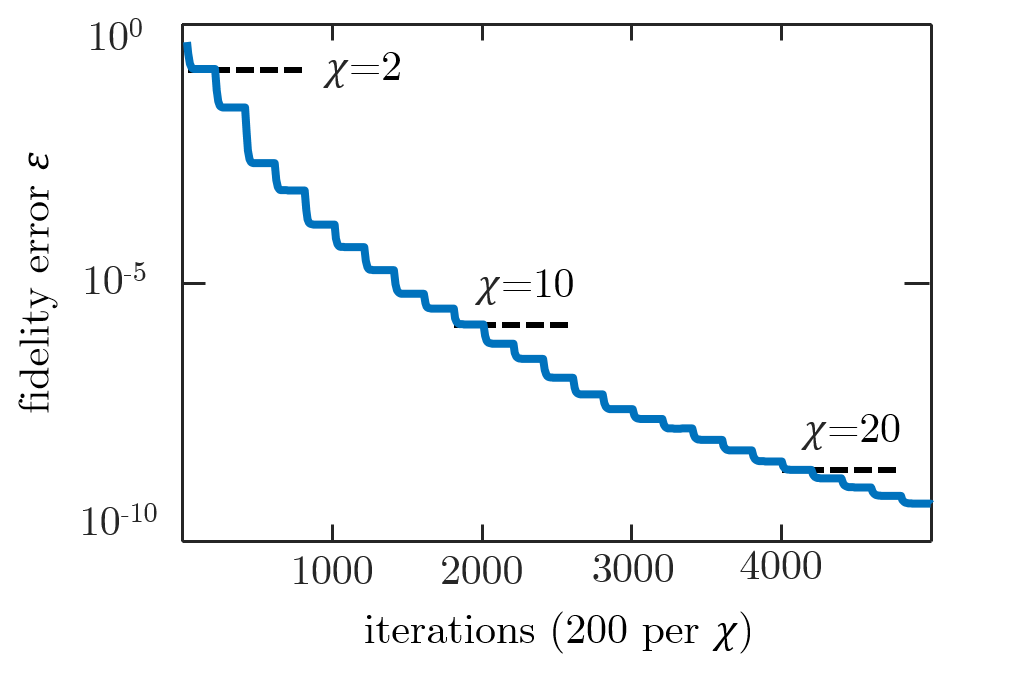}
\caption{Convergence sequence for the ground state of a random inhomogeneous Hamiltonian with $d=2$, $l=3$, and open boundary conditions on a $N=16$ site lattice using a 50\% sample rate. Several bond dimension $\chi$ are noted and labeled accordingly, corresponding to convergences in the specified $\chi$-value. When contracting a tensor network, $\chi$ reflects the number of singular values kept across  The exponential decay in fidelity error $\varepsilon$ also mirrors the singular value spectra corresponding to the random Hamiltonians seen in Fig. \ref{fig:svals_example}.}
\label{fig:convergence_and_svals}
\end{center}
\end{figure}

Fig. \ref{fig:convergence_and_svals} demonstrates the convergence properties of our approach using the ground state of a randomly generated Hamiltonian. The jagged presentation of the convergence plot reflects convergences in each bond dimension, where the $\chi$-level convergence is used as the initial state for the next highest bond dimension ($\chi+1$). This process repeats until there is negligible change between bond dimensions, i.e. the state is complete. Comparing Fig. \ref{fig:convergence_and_svals} to the dashed lines in Fig. \ref{fig:svals_example} (singular value spectra for two random Hamiltonians), we see a similar exponential decay pattern, indicating that our approach is adding refinement to the solution at a rate consistent with the singular values. 
 
\begin{figure}[!t!b]
\begin{center}
\includegraphics[width=8.5cm]{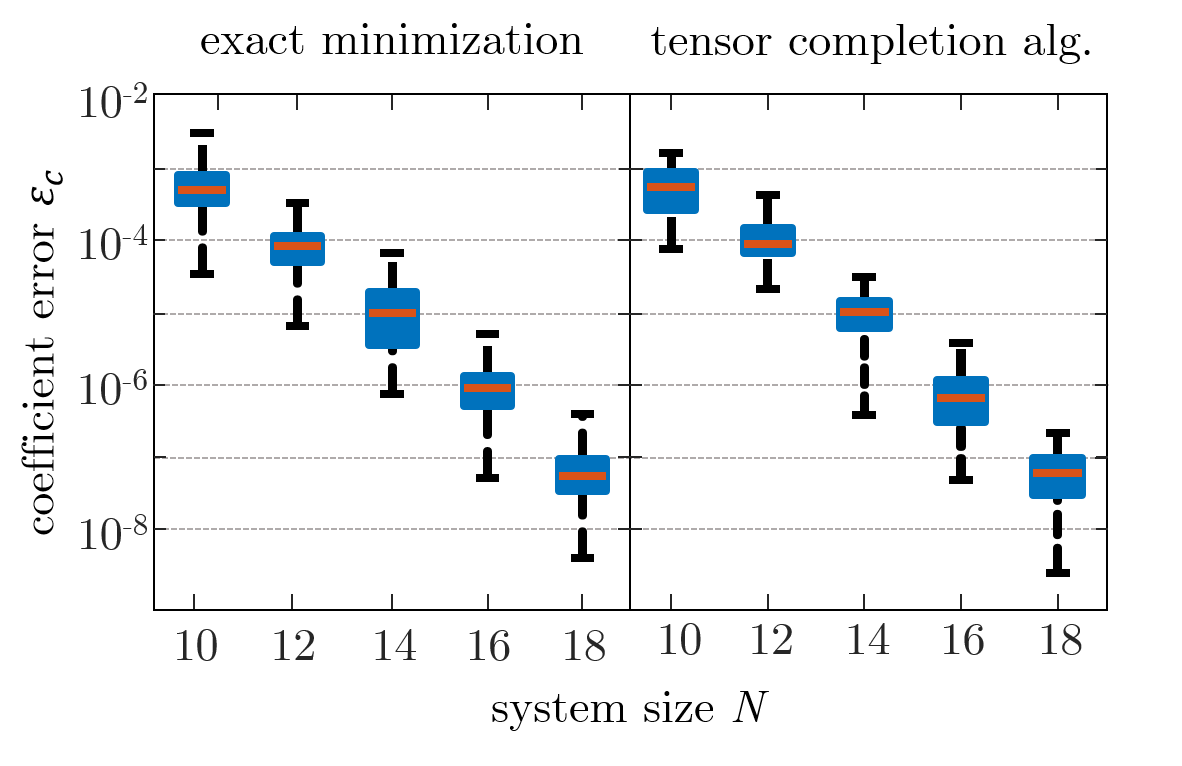}
\caption{Comparing coefficient errors $\varepsilon_{c}$ between exact entanglement minimization over entire parameter space (left) as performed in Fig. \ref{fig:alpha_plot} and our tensor completion algorithm (right). For consistency across system size, the same randomly generated homogeneous local operator of local dimension $d=2$ is used for all $N$. 
At each $N$, 50 coefficients are randomly chosen from the corresponding ground state; for exact minimization, we solve for the minimum $S_{1/2}$ coefficient value of each coefficient individually.  
Each whisker plot includes 50 randomly selected coefficients at each $N$ and calculate $\varepsilon_{c}$ (left) for each coefficient individually using exact minimization and (right) for all coefficients simultaneously using our tensor completion algorithm. The 50 resulting $\varepsilon_{c}$ are plotted via whisker plots for both cases at each $N$. The dotted grey line reflects the median result for the exact minimization approach (left), demonstrating that our algorithm reliably obtains the optimal minimal entanglement solution for each coefficient.}
\label{fig:exact_vs_alg}
\end{center}
\end{figure}

Having shown in Sec. III that the minimal entanglement solution converges to the exact ground state exponentially quickly in $N$, we evaluate the degree to which our tensor-based completion algorithm achieves the same result. In Fig. \ref{fig:exact_vs_alg} coefficient errors $\varepsilon_{c}$ from the exact minimization approach deployed in Sec. III are compared to $\varepsilon_{c}$ obtained from our tensor completion algorithm. In the former case, we calculate $\varepsilon_{c}$ for each unsampled coefficient $c \in S'$ individually, as performed in Fig. \ref{fig:alpha_plot} and Table \ref{table:alpha_table}, whereas the completion algorithm solves for all $c \in S'$ simultaneously. To enforce consistency across $N$, we use the same randomly generated homogeneous operator with periodic boundary conditions, $d=2$, and $l=2$ for all $N$. At each $N$, $d^N - 50$ random samples are taken, giving $S'$ a cardinality of 50. As shown in Fig. \ref{fig:exact_vs_alg}, both approaches yield approximately the same result across all $c \in S'$ at each $N$. We found this to be true in general for all locally defined ground states, provided that the sample rate is sufficiently high. Therefore we conclude that a sufficiently large random sample of coefficients $S$ for a given $N$, our tensor completion algorithm returns the optimal minimal entanglement solution. The minimum $S$ required for this to hold depends on many factors, including the inherent randomness of randomly generated operators, and is explored below.

The result in Fig. \ref{fig:exact_vs_alg} utilizes sample rates of 95\% for $N=10$ and higher still for the larger systems, but our tensor completion algorithm converges to the minimal entanglement solution across a robust range of sample rates, depending on system parameters and the randomness of randomly generated operators. The precise limit of the sample rate required for the minimal entanglement completion depends on the many parameters of a locally defined ground state (homogeneity, boundary conditions, local dimension, interaction length, number of lattice sites, etc.) as well as the randomness inherent in randomly generated operators. As such, establishing specific limits for this is beyond the scope of this paper. 

\begin{figure}[!t!b]
\begin{center}
\includegraphics[width=8.7cm]{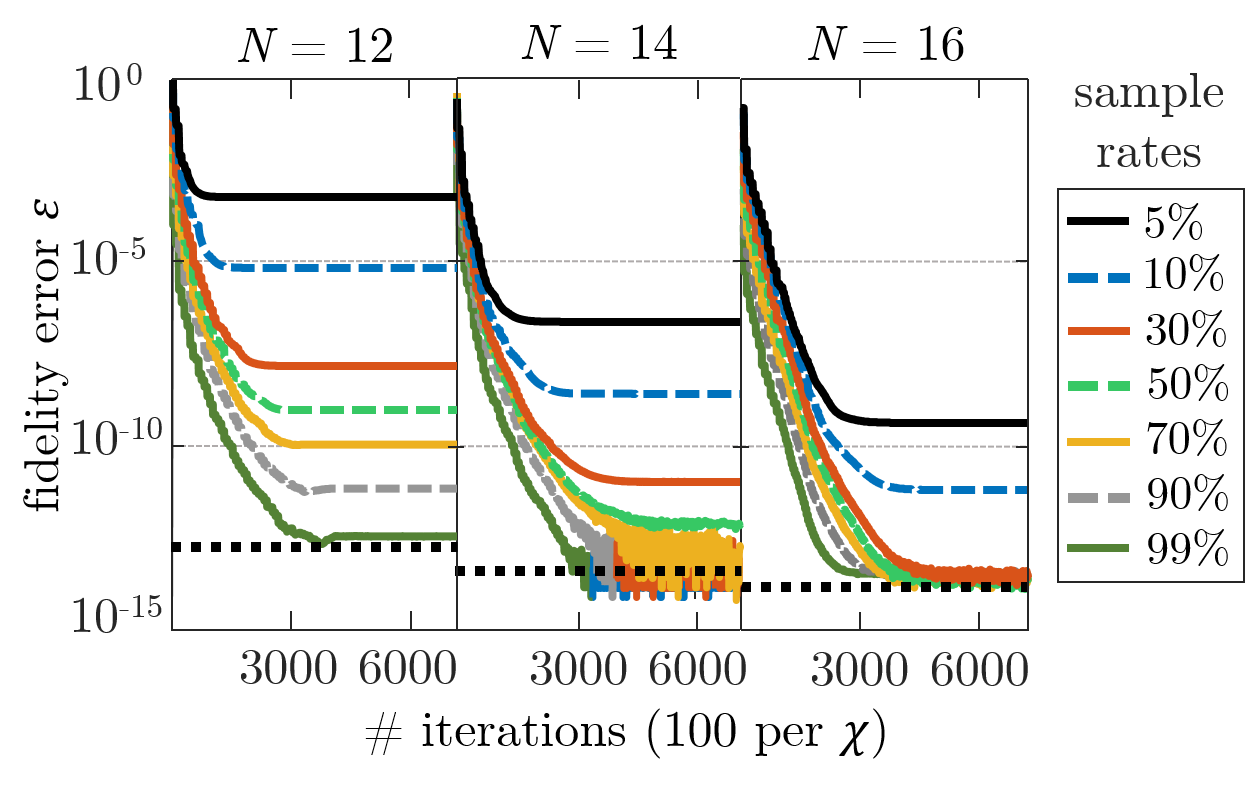}
\caption{Convergence sequences for the completion of three random local ground states using various sample rates, where system size is varied from $N=12$ (left) to $N=16$ (right). Interaction length $l=2$ and local dimension $d=2$ are used in each case. The dotted black line at the bottom of each plot represents the minimally entangled solution, which reflects the theoretical maximum accuracy of an $S_{1/2}$ minimization-based completion algorithm and is extrapolated from the exact $S_{1/2}$ minimization technique in Sec. III.}
\label{fig:convergences}
\end{center}
\end{figure}

Increasing system size with all else equal, we find that the minimal entanglement solution becomes available to an ever larger sample rate domain. Fig. \ref{fig:convergences} includes convergence sequences for a range of sample rates across three lattice sizes, $N=12$ (left), $N=14$ (center), and $N=16$ (right). The $N=16$ plot, possessing the largest ratio of system size to interaction length and local dimension, returns the minimal entanglement solution (indicated by the dotted black line) for all sample rates $\geq$ 30\%. As $N$ is lowered to $N=12$ (right), however, a much larger sample rate $\geq 99\%$ is required to obtain the equivalent minimal entanglement solution for the smaller sized system. The $N=14$ result falls somewhere in the middle, with a sample rate $\geq$ 70\% required for the optimal result. All plots demonstrate that beyond the sample rate required for a minimal entanglement solution, the completion algorithm still returns a completed state that maintains high levels of accuracy. Insofar as this is the case, completion routines can still be useful when operating on local ground states, even if the sample rate is too low to obtain the optimal (minimum $S_{1/2}$ entanglement) solution. The $N=12$ plot also demonstrates the largest deviation from established results from matrix completion, where low-rank matrices can be completed with zero error as long as the sample rate is high enough. Overall, Fig. \ref{fig:convergences} suggests that $N$ needs to be roughly 1 order of magnitude larger than $d$ and $l$ for the minimal entanglement solution to be accessible to a meaningful range of sample rates (e.g., sample rates $\in [30\% : 99\%]$ for the random $N=16$ Hamiltonian featured in the figure).

\begin{figure}[!t!b]
\begin{center}
\includegraphics[width=8.5cm]{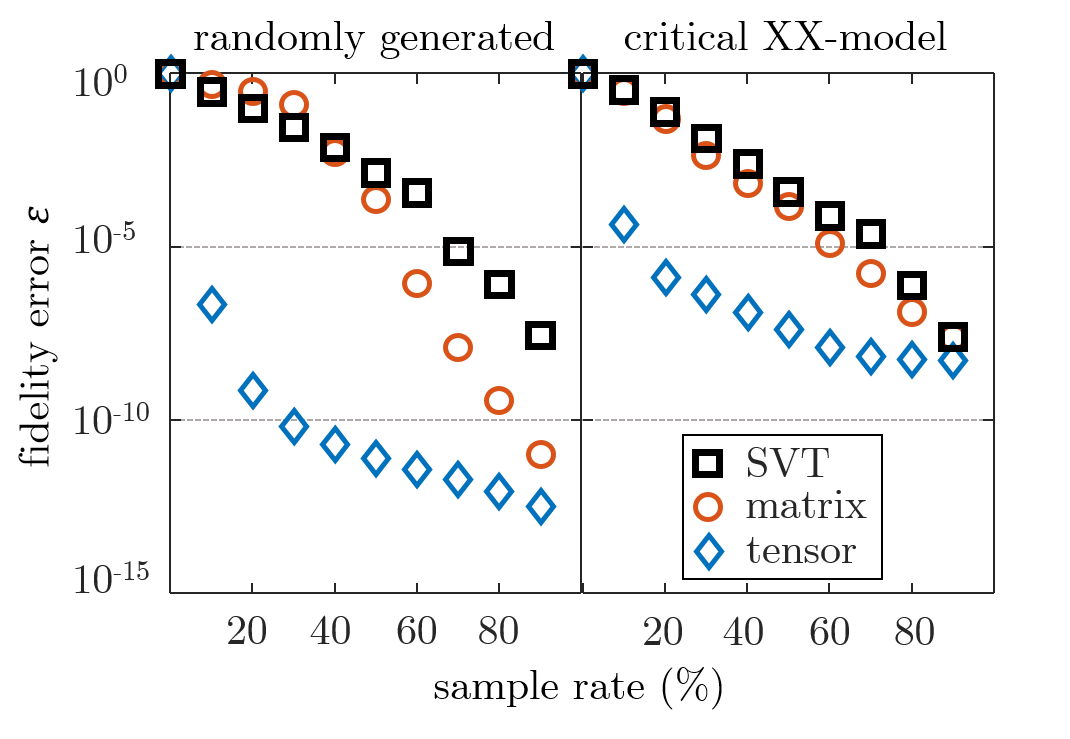}
\caption{Comparing completion accuracy from partial samples at various sample rates using three different methods: SVT (singular value thresholding from Cand{\'e}s et al.)\cite{MCM3}, matrix (the algorithmic scaffolding from Sec. IV using only a central bipartition; i.e., treating the wavefunction as a matrix and gradually ramping up the bond dimension per the algorithmic scaffolding), and tensor (the full randomized tensor tree approach outlined in Sec. IV). The left plot features results for ground states of randomly generated homogeneous (translationally invariant) complex Hamiltonians of interaction length $l=2$ and local dimension $d=2$, and right plot contains results for ground states of the critical XX-model Hamiltonian. All data were obtained from a $N=14$ site 1D lattice with periodic boundary conditions. The same partial samples were used as the starting point for each method to enforce consistent initial conditions.}
\label{fig:method_comp}
\end{center}
\end{figure}

A priori, it wasn't clear that leveraging all possible bipartitions of a ground state would necessarily yield a more accurate ground state completion algorithm as compared to traditional matrix-based completion approaches, or tensor completion algorithms where the tensor is treated with fixed dimensionality. However, we find that a tensor-based approach that iterates over all bipartitions is superior across the entirety of the sample rate domain. In Fig. \ref{fig:method_comp}, we compare our completion algorithm to a gold-standard in noiseless matrix completion, the singular value thresholding (SVT) routine from Cand{\'e}s et al. We also include a matrix-based version of our algorithm, which leverages the algorithmic scaffolding introduced in Sec. IV, but only performs T-SVDs across a central bipartition (thereby treating the ground state as a matrix with fixed dimensions). Interestingly, this approach performs better than the SVT algorithm across the majority of the sample rate domain. This improvement is indicative that our sequential ramping up of bond dimension confers an accuracy advantage in the context of wavefunctions, and potentially for other data that have singular value spectra similar to those of ground states.

\begin{figure}[!t!b]
\begin{center}
\includegraphics[width=8.7cm]{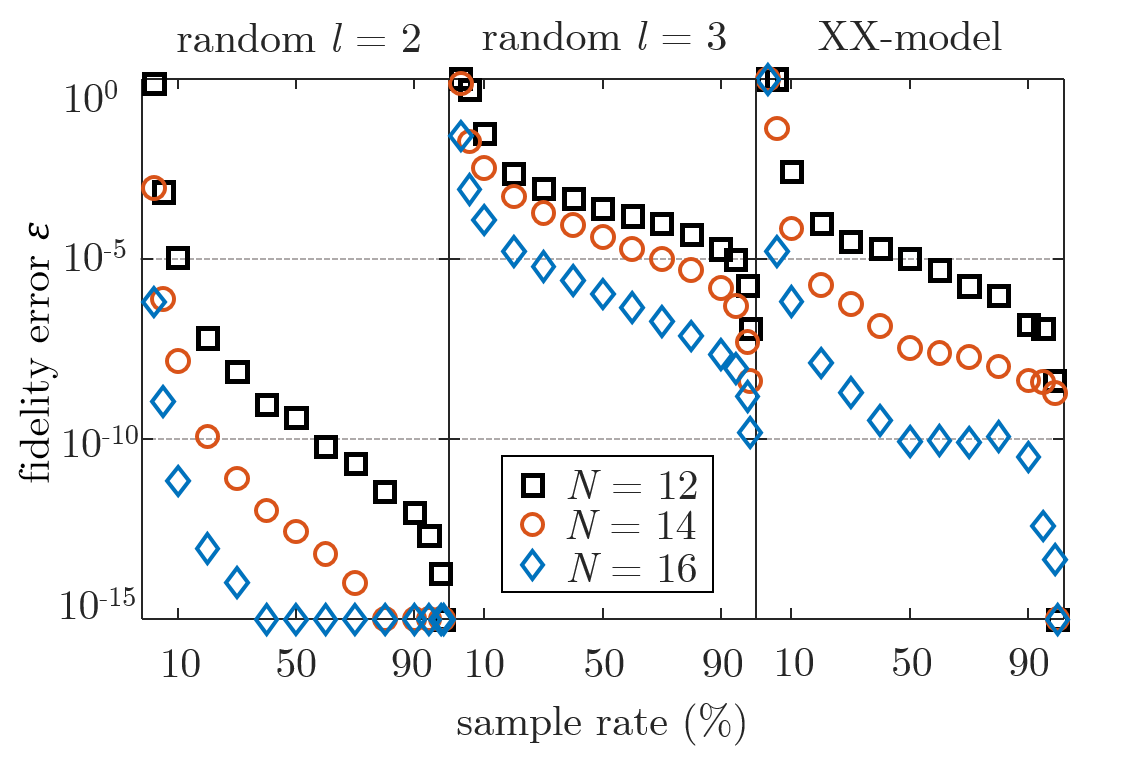}
\caption{Median fidelity error measured over 20 trials per data point for three different types of systems and across three differently sized lattices N=\{12, 14, 16\}. The two random systems (left and center) correspond to randomly generated inhomogeneous complex Hamiltonians with periodic boundary conditions and local dimension $d=2$, only differing in their respective interaction lengths $l$ labeled at the top. Results for the critical XX-model with open boundary conditions are plotted in the right pane. The complete sample rate domain included in each graph (in percentage \%): \{2, 5, 10, 20, 30, 40, 50, 60, 70, 80, 90, 95, 99, 99.9\}.}
\label{fig:alg_performance}
\end{center}
\end{figure}

In Fig. \ref{fig:alg_performance}, we include several performance plots for our tensor completion algorithm that encompass three types of local ground states on three different lattice sizes. Of note is the significant difference in fidelity error between the ground states of randomly generated Hamiltonians of interaction length $l=2$ and $l=3$ (left and center, respectively). In the case of the $N=16$ site lattice and $l=2$, the limits of numerical precision are reached in fidelity error for all sample rates $\geq 40\%$, whereas for $l=3$ these limits are never met. This is consistent with our earlier finding of the theoretical maximum accuracy conferred by the minimal entanglement solution of unsampled coefficients (see Sec. III). Specifically in Fig. \ref{fig:alpha_plot}, for $N=16$ and $d=2$, the difference in coefficient error $\varepsilon_{c}$ of the minimal $S_{1/2}$ entanglement solution for $l=2$ and $l=3$ spans roughly 6 orders of magnitude. From sample rates of ~10\% to ~90\% in Fig. \ref{fig:alg_performance}, we see a similar spread of roughly 6 orders of magnitude in fidelity error between the $l=2$ and $l=3$ cases.

\section{Conclusion and Outlook}
We have shown that ground states of local systems on 1D lattices are well approximated by minimal entanglement completions of their partial samples. More precisely, the error $\varepsilon$ of a minimal entanglement completion, as measured by the fidelity, was numerically demonstrated to converge to zero exponentially quickly in the system size, $\varepsilon \approx \exp(-\alpha N)$, with coefficient $\alpha$ inversely related to the amount of entanglement in the system. This result contrasts with established results from matrix completion, where it is known that low-rank matrices can be reconstructed with essentially zero error provided that the sample size of matrix entries is large enough. That a minimal entanglement completion of a ground state can have significant error, even if the completion is optimal, can be understood given that ground states, which follow the boundary law for entanglement entropy scaling, are not necessarily low-rank. While the singular value spectra of ground state across local bi-partitions do decay sharply, it can still be the case that the number of non-negligibly small singular values is still relatively large.       

We have also proposed a new algorithm for tensor completion based on iterative decomposition of the trial wavefunction into (one of many possible choices of) tensor network states. This demonstrates a novel application of tensor network methods in physics beyond their usual application: just as tensor networks offer can an efficient representation of quantum ground states, their structure can equally be exploited to fill-in missing information from quantum ground states. Our proposed algorithm, although surprisingly simple, was numerically demonstrated to produce close-to optimal completions, given that the errors in the recovered elements matched those given from a brute-force search over the full parameter space. An appealing feature of our tensor completion algorithm is that it is shown to work even when the tensors/matrices under consideration are not low rank; in many cases the ground state wavefunctions were completed to high accuracy even though they were full-rank (across any local bipartition) to within the desired accuracy. We attribute the success of our proposed algorithm in this setting, where many previous algorithms for matrix/tensor completion struggle, to our iterative ramping-up of the tensor bond-dimension in conjunction with the form of the singular value spectra which, although not low-rank, still decay sharply. It seems likely that our proposed algorithm for tensor completion could find useful applications towards other problems. 

Although we were unable to fully characterise the scaling of the minimum sample size needed for successful completion as a function of system size $N$, we did show that the sample size was far below the theoretical limit for the comparable matrix completion problem given from an equally-sized bipartition of the 1D spin chain. In other words the ground states of systems with local interactions, which are known to obey the boundary law for entanglement, possess a remarkably low information density; they can be completed from fewer samples than is possible for any comparatively sized matrix. This result is attributed to the fact that ground state wavefunctions are simultaneously low-entanglement across all local bipartitions, not just a single bipartition. The success of our completion algorithm also further illustrates the extent to which ground state wavefunctions are highly atypical as compared to random states in the Hilbert Space. We speculate that it could ultimately be possible to reconstruct the ground state on an $N$-site lattice with $\textrm{poly}(N)$ samples of its wavefunction amplitudes, although cannot provide a conclusive evidence to this point without a more sophisticated numerical investigation.  

Finally, although the present work was not focused on the application to improve simulation algorithms for quantum many-body systems, we remark that our proposed tensor network completion algorithms could be useful for sampling-based numerical simulation algorithms (e.g. quantum Monte-carlo) where they could allow for improved accuracy from smaller sample sizes. Similarly, they might also be employed to improve sampling-based methods for the contraction of tensor networks. These applications remain an interesting avenue of exploration for future work. 



\newpage
\begin{thebibliography}{99}



\bibitem{TN1}
J. I. Cirac and F. Verstraete, {\it Renormalization and tensor product states in spin chains and lattices}, J. Phys. A: Math. Theor. 42, 504004 (2009).

\bibitem{TN2}
R. Orus, {\it A practical introduction to tensor networks: Matrix product states and projected entangled pair states}, Ann. Phys. 349, 117 (2014).

\bibitem{TN3}
J. C. Bridgeman and C. T. Chubb, {\it Hand-waving and Interpretive Dance: An Introductory Course on Tensor Networks}, J. Phys. A: Math. Theor. 50, 223001 (2017).


\bibitem{QC1}
G. K. L. Chan and S. Sharma, {\it The density matrix renormalization group in quantum chemistry}, Annu. Rev. Phys. Chem. 62, 465 (2011). 

\bibitem{QC2}
N. Nakatani and G. K.-L. Chan, {\it Efficient tree tensor network states (TTNS) for quantum chemistry: generalizations of the density matrix renormalization group algorithm}, J. Chem. Phys. 138, 134113 (2013).


\bibitem{ML1}
A. Anandkumar, R. Ge, D. Hsu, S. M. Kakade and M. Telgarsky, {\it Tensor decompositions for learning latent variable models}, Journal of Machine Learning Research 15, 2773–2832 (2014).

\bibitem{ML2}
A. Novikov, D. Podoprikhin, A. Osokin and D. Vetrov, {\it Tensorizing neural networks}, arxiv:1509.06569 (2015).

\bibitem{ML3}
E. M. Stoudenmire and D. J. Schwab, {\it Supervised learning with tensor networks}, Advances In Neural Information Processing Systems 29, pp. 4799–4807 (2016).


\bibitem{Holo1}
B. Swingle, {\it Entanglement renormalization and holography}, Phys. Rev. D 86, 065007 (2012).

\bibitem{Holo2}
B. Swingle, {\it Constructing holographic spacetimes using entanglement renormalization}, arXiv:1209.3304 (2012).

\bibitem{Holo3}
P. Hayden, S. Nezami, X. L. Qi, N. Thomas, M. Walter and Z. Yang, {\it Holographic duality from random tensor networks}, arXiv:1601.01694 (2016).

\bibitem{Holo4}
G. Evenbly, {\it Hyperinvariant tensor networks and holography}, Phys. Rev. Lett. 119, 141602 (2017).


\bibitem{DMRG1}
S. R. White, {\it Density matrix formulation for quantum renormalization groups}, Phys. Rev. Lett. 69, 2863 (1992).

\bibitem{DMRG2}
S. R. White, {\it Density-matrix algorithms for quantum renormalization groups}, Phys. Rev. B 48, 10345 (1993). 

\bibitem{DMRG3}
U. Schollwoeck, {\it The density-matrix renormalization group}, Rev. Mod. Phys. 77, 259 (2005).


\bibitem{MPS1}
M. Fannes, B. Nachtergaele, and R. F. Werner, {\it Finitely correlated states on quantum spin chains}, Commun. Math. Phys. 144, 443 (1992).

\bibitem{MPS2}
S. Ostlund and S. Rommer, {\it Thermodynamic limit of density matrix renormalization}, Phys. Rev. Lett. 75, 3537 (1995).

\bibitem{MPS3}
G. Vidal, {\it Efficient classical simulation of slightly entangled quantum computations}, Phys. Rev. Lett. 91, 147902 (2003).


\bibitem{PEPS1} 
F. Verstraete and J. I. Cirac, {\it Renormalization algorithms for quantum-many-body systems in two and higher dimensions}, arXiv:cond-mat/0407066.

\bibitem{PEPS2} 
F. Verstraete, J.I. Cirac, and V. Murg, {\it Matrix product states, projected entangled pair states, and variational renormalization group methods for quantum spin systems}, Adv. Phys. 57, 143 (2008).

\bibitem{PEPS3}
J. Jordan, R. Orus, G. Vidal, F. Verstraete, and J. I. Cirac, {\it Classical simulation of infinite-size quantum lattice systems in two spatial dimensions}, Phys. Rev. Lett. 101, 250602 (2008).

\bibitem{PEPS4}
H. N. Phien, J. A. Bengua, H. D. Tuan, P. Corboz and Roman Orus, {\it The iPEPS algorithm, improved: fast full update and gauge fixing}, Phys. Rev. B 92, 035142 (2015).


\bibitem{TTN1}
Y. Shi, L. Duan and G. Vidal, {\it Classical simulation of quantum many-body systems with a tree tensor network}, Phys. Rev. A 74, 022320 (2006).

\bibitem{TTN2}
L. Tagliacozzo, G. Evenbly and G. Vidal, {\it Simulation of two-dimensional quantum systems using a tree tensor network that exploits the entropic area law}, Phys. Rev. B 80, 235127 (2009).


\bibitem{MERA1}
G. Vidal, {\it A class of quantum many-body states that can be efficiently simulated}, Phys. Rev. Lett. 101, 110501 (2008).

\bibitem{MERA2}
L. Cincio, J. Dziarmaga, and M. M. Rams, {\it Multiscale entanglement renormalization ansatz in two dimensions: quantum Ising model}, Phys. Rev. Lett. 100, 240603 (2008).

\bibitem{MERA3} 
G. Evenbly and G. Vidal, {\it Entanglement renormalization in two spatial dimensions}, Phys. Rev. Lett. 102, 180406 (2009).

\bibitem{MERA4} 
G. Evenbly and G. Vidal, {\it Quantum criticality with the multi-scale entanglement renormalization ansatz}, Chapter 4 in \textit{Strongly Correlated Systems: Numerical Methods}, edited by A. Avella and F. Mancini (Springer Series in Solid-State Sciences, Vol. 176 2013).


\bibitem{Ent1}
J. Eisert, M. Cramer, and M. Plenio, {\it Area laws for the entanglement entropy - a review}, Reviews of Modern Physics, 82, pp. 277-306 (2010).

\bibitem{Ent2}
M B Hastings, {\it An area law for one-dimensional quantum systems}, J. Stat. Mech. P08024 (2007)

\bibitem{Ent3}
J. I. Latorre, E. Rico, and G. Vidal, {\it Ground state entanglement  quantum spin chains}, Quantum Inf. Comput., 4, pp. 48-92 (2004).

\bibitem{Ent4}
J. Eisert, {\it Entanglement and tensor network states}, arXiv quant-ph, 1308.3318 (2013).

\bibitem{Ent5}
S. Irani, {\it Ground state entanglement in one-dimensional translationally invariant quantum systems}, J. Math. Phys., 51, 022101 (2010).

\bibitem{Ent6}
J. Ren, W. You, and X. Wang, {\it Entanglement and correlations in a one-dimensional quantum spin $\frac{1}{2}$ chain with anisotropic power-law long-range interactions}, Phys. Rev. B. 101, 094410 (2020).


\bibitem{MCM1}
E. Cand{\`e}s and B. Recht, {\it Exact matrix completion via convex optimization}, Found. Comput. Math., 9, 717 (2009).

\bibitem{MCM2}
E. Cand{\`e}s and T. Tao, {\it The power of convex relaxation: near-optimal matrix completion}, IEEE Trans. on Inform. Theory, 56, 5 (2010). 

\bibitem{MCM3}
J. Cai, E. Cand{\`e}s, and Z. Shen, {\it A singular value thresholding algorithm for matrix completion}, SIAM Journal on Optim., 20, 4 (2010).

\bibitem{MCM4}
B. Recht, {\it A simpler approach to matrix completion}, Journal of Machine Learning Research, 12, pp. 3413-3430 (2011).

\bibitem{MCM5}
R. H. Keshavan, A. Montanari, and S. Oh, {\it Matrix completion from a few entries}, IEEE Transactions on Information Theory, 56, 6, pp. 2980-2998 (2010).

\bibitem{MCM6}
B. Vandereycken, {\it Low-rank matrix completion by Riemannian optimization}, SIAM J. Optim., 23, 2 (2013).

\bibitem{MCM7}
J. R. Johnson, {\it Matrix completion problems: a survey}, Matrix Theory and Applications, 40, pp. 171-198 (1990).

\bibitem{MCM8}
L. T. Nguyen, J. Kim, and B. Shim, {\it Low-rank matrix completion: a contemporary survey}, IEEE Access, 7, pp. 94215-94237 (2019).

\bibitem{MCM9}
R. Ge, J. Lee, and T. Ma, {\it Matrix completion has no spurious local minimum}, NIPS (2016).

\bibitem{MCM10}
M. Fazel, H. Hindi, and S. Boyd, {\it Rank minimization and applications in system theory}, Proceedings of the 2004 American Control Conference, 4, pp. 3273-3278 (2004). 94215-94237 (2019).

\bibitem{MCM11}
S. Ma, D. Goldfarb, and L. Chen, {\it Fixed point and Bregman iterative methods for matrix rank minimization}, Mathematical Programming, 128, pp. 321-353 (2011).


\bibitem{MCA10}
C. A. Gomez-Uribe and N. Hunt, {\it The Netflix recommender system: algorithms, business value, and innovation}, Association for Computing Machinery, 6, 4 (2016).

\bibitem{MCA1}
A. Ramlatchan, M. Yang, Q. Liu, M. Li, J. Wang, and Y. Li, {\it A survey of matrix completion methods for recommendation systems}, Big Data Mining and Analytics, 1, 4 (2018).

\bibitem{MCA2}
E. Acar, D. Dunlavy, and T. Kolda, {\it Link prediction on evolving data using matrix and tensor factorizations}, 2009 IEEE International Conference on Data Mining Workshops, pp. 262-269 (2009).

\bibitem{MCA3}
O. Troyanskaya, M. Cantor, G. Sherlock, P. Brown, T. Hastie, R. Tibshirani, D. Botstein, R. Altman, {Missing value estimation methods for DNA microarrays}, Bioinformatics, 17, 6 (2001).

\bibitem{MCA4}
E. Chi et al., {\it Genotype imputation via matrix completion}, Genome Research, 23, pp. 509-518 (2013). 

\bibitem{MCA5}
E. Cand{\`e}s, Y. Eldar, T. Strohmer, and V. Voroninski, {Phase retrieval via matrix completion}, SIAM J. on Imaging Sciences, 6, 1 (2011).

\bibitem{MCA6}
H. Xue, S. Zhang, and D. Cai, {Depth image inpainting: improving low rank matrix completion with low gradient regularization}," in IEEE Transactions on Image Processing, 26, 9 (2017).

\bibitem{Sens1}
E. Cand{\`e}s and J. Romberg, {\it Sparsity and incoherence in compressive sampling}, Inverse Problems, 23, 3, pp. 969-985 (2007).

\bibitem{Sens2}
D. Gross et al., {\it Quantum state tomography via compressed sensing}, Phys. Rev. Lett., 105, 15 (2010).

\bibitem{Sens3}
M. Kliesch, R. Kueng, J. Eisert, and D. Gross, {\it Improving compressed sensing with the diamond norm}, IEEE Trans. Inf. Theory, 62, 7445  (2016).


\bibitem{TCM1}
T. Kolda and B. Bader, {\it Tensor Decompositions and Applications}, SIAM Review, 51, 3 (2009).

\bibitem{TCM2}
M. Ashrapijuo and X. Wang, {\it Fundamental conditions for low-CP-rank tensor completion}, Journal of Machine Learning Research, 18, 63 (2017).

\bibitem{TCM3}
A. Montanari and N. Sun, {\it Spectral algorithms for tensor completion}, Comm. on Pure and Applied Math., 71, 11 (2018).

\bibitem{TCM4}
A. Liu and A. Moitra, {\it Tensor Completion Made Practical}, arXiv:2006.03134 (2020).

\bibitem{TCM5}
C. Ko et al., {\it Fast and accurate tensor completion with total variation regularized tensor trains}, IEEE Trans. on Image Proc., 29, pp. 6918-6931 (2020).


\bibitem{TCA1}
J. Liu, P. Musialski, P. Wonka and J. Ye, {\it Tensor Completion for Estimating Missing Values in Visual Data}, IEEE Transactions on Pattern Analysis and Machine Intelligence, 35, 1 (2013).

\bibitem{TCA2}
M. Signoretto, R. Van de Plas, B. De Moor, and J. A. K. Suykens, {\it Tensor versus matrix completion: a comparison with applications to spectral data}, IEEE Signal Processing Letters, 18, 7 (2011).

\bibitem{TCA3}
Q. Song, H. Ge, J. Caverlee, and X. Hu, {\it Tensor completion algorithms in big data analytics}, ACM Trans. Knowl. Discov. Data, 13, 1 (2019)


\bibitem{MISC1}
J.C. Lagarias et al., {\it Convergence properties of the Nelder-Mead Simplex Method in Low Dimensions}, SIAM Journal of Optim., 9, 1, pp. 112-147 (1998).


\end{thebibliography}
\end{document}